\documentclass[12pt]{iopart}

\usepackage{graphicx}% Include figure files
\usepackage{dcolumn}% Align table columns on decimal point
\usepackage{bm}% bold math
\usepackage{epstopdf}
\usepackage[utf8]{inputenc}
\usepackage{amsmath}

\usepackage[unicode=true,pdfusetitle,
 bookmarks=true,bookmarksnumbered=false,bookmarksopen=false,
 breaklinks=false,pdfborder={0 0 0},backref=false,colorlinks=true,citecolor=blue,urlcolor=violet 
]{hyperref}
\usepackage{xcolor}
\usepackage{graphicx}
\usepackage{iopams}  
\begin{document}

\title[1-d Ising model using Kronecker sum and Kronecker product]{1-d Ising model using Kronecker sum and Kronecker product}

\author{Sourabh Magare $^{1}$ $^{2}$ , Abhinash Kumar Roy$^{1}$, Varun Srivastava$^{1}$}

\vspace{10pt}
%\affiliation{ Department of Physical Sciences, Indian Institute of Science Education and Research– Kolkata, Mohanpur Campus, Nadia 741246, West Bengal, India}

\address{%
$^{1}$ \quad Indian Institute of Science Education and Research Kolkata, Mohanpur, West Bengal, 741246, India.\\
$^{2}$ \quad Inter University Centre for Astronomy and Astrophysics, Pune 411007.}

\vspace{10pt}
 \ead{\href{mailto:sourabh.magare@gmail.com}{sourabh.magare@iucaa.in}, \href{mailto:akr16ms137@iiserkol.ac.in}{akr16ms137@iiserkol.ac.in},\href{mailto:vs16ms109@iiserkol.ac.in} {vs16ms109@iiserkol.ac.in}}
%\begin{indented}
%\item[]August 2017
%\end{indented}

\begin{abstract}

Calculations in Ising model can be cumbersome and non-intuitive. Here we provide a formulation that addresses these issues for 1-D scenario. We represent the microstates of spin interactions as a diagonal matrix. This is done using two operations: Kronecker sum  and Kronecker product. The calculations thus become simple matter of manipulating diagonal matrices. We address the following problems in this work: spins in the magnetic field, open-chain 1-D Ising model, closed-chain 1-D Ising model, 1-D Ising model in an external magnetic field. We believe that this representation will help provide students as well as experts with a simple yet powerful technique to carry out calculations in this model.
\end{abstract}

%\maketitle

\section{Introduction}
It is remarkable that many complicated statistical systems can be studied using relatively simple mathematical models involving lattice arrangements of molecules and considering the nearest-neighbour interactions \cite{RevModPhys.39.883}.One such model that has seen wide range of applications is the Ising Model \cite{1925ZPhy...31..253I} proposed in 1925 by Wilhelm Lenz and solved for 1-D spin lattice by Ernst Ising as a part of his doctoral thesis \cite{RevModPhys.39.883}.    

It is ironic, that after showing there could be no phase transitions in 1-D systems at $T \ne 0$ and erroneously concluding that this was true for higher dimensional systems as well, Ising gave up study of this model and realized much later that his name had become immortal because of it \cite{10.2307/24097007}. In today's world, Ising model has found wide ranging applications in various fields. It is one of the simplest models that shows phase transitions of statistical systems in higher dimensions. Building on the works of H.A Kramers and G. H. Wannier \cite{PhysRev.60.252}, Lars Onsager gave an exact solution for 2-D Ising model which showed phase transitions and is considered one of the landmarks in theoretical physics \cite{PhysRev.65.117}. This model has also played a crucial role in studying alloys \cite{10.2307/96591,1935RSPSA.150..552B}, spin glasses \cite{Kawamura_2010}, in neuroscience \cite{Hopfield2554,Schneidman06weakpairwise}  and even in modelling financial markets \cite{2015JPhCS.574a2149T, doi:10.1119/1.2779882, doi:10.1142/S0129183101001845} and studying epidemics and pandemics with reference to recent COVID-19 outbreak \cite{2021PhyA..57325963M}. Heisenberg's model, which was inspired by Ising model \cite{RevModPhys.39.883} is finding wide scale applications in quantum information and quantum computing \cite{2004PhRvA..70e2322L,2016NatCo...713070M,PhysRevA.70.032313}.

Even though the model was proposed almost a century ago, it is clear that its applications are still being found in many important areas. At its heart, calculations in Ising model involve counting various microstates of the system. This procedure then helps us to calculate the partition function which embeds information of the macroscopic properties of the system. The most
widely taught method to solve the Ising model exactly is
the transfer matrix method \cite{baxter2007exactly}.
Mathematically, solving Ising model is a combinatorial problem and people have given purely combinatorial techniques for eg. Kac and Ward's work \cite{PhysRev.88.1332} using combinatorics to yield the partition function of 2-D Ising model and Feynman's contribution towards this work \cite{costa2003f}.Recently, numerical techniques have been applied to study higher dimensional Ising models \cite{PARK2022107325,1993JSP, 10.1214/09-AOS691,PhysRevLett.62.361,2016} and studying long range interactions in Ising chains; see \cite{Nagle_1970,1990,PhysRevB.52.3034} and references therein.

Our aim here is to provide a method that is more physically intuitive and less cumbersome for the 1-D scenario. There exist numerous approach to exactly solve the Ising model (for some recent works see  \cite{Seth_2016,Wang_2019}) for various geometries and configurations \cite{1978}. However, we believe that the method presented in this article provides a simpler and yet powerful approach using the operations of Kronecker sum and Kronecker product. 

The paper is organized as follows: In section II, we review the definition and properties of  Kronecker sum and Kronecker product and discuss them in the context of diagonal matrices. In section III, we provide a detailed prescription to obtain partition function for spin interaction Hamiltonians using spin-1/2 particles in the absence of external magnetic field employing the Kronecker product and sum operations. In section IV we develop our approach for non-interacting spins in the presence of an external magnetic field. Section V is dedicated to solving the 1-D Ising model for open/close chains in absence and presence of external magnetic fields.

\section{Preleminaries}
In this section, we review the definitions and properties of Kronecker sum and Kronecker product. Kronecker product ($\otimes$) operation (also known as tensor product) is defined as following \cite{britanak2010discrete}. If S is a $m\times n$ matrix and T is a $p \times q$ matrix then the Kronecker product is a $pm \times nq$ matrix:   
\begin{equation}
    S \otimes T = \begin{bmatrix}
s_{11}& . & . & s_{1n}\\
.& . & . & .\\
.& . & . & .\\
s_{m1}& . & . & s_{mn}\\
\end{bmatrix}
\otimes T = \begin{bmatrix}
s_{11} T& . & . & s_{1n} T\\
.& . & . & .\\
.& . & . & .\\
s_{m1} T& . & . & s_{mn} T\\
\end{bmatrix} .
\end{equation}
Kronecker product is an operation on matrices of arbitrary sizes. It is important to note that Kronecker product is associative however, non-commutative. Moreover, it is distributive over the usual addition, i.e., $(A + B)\otimes C = A \otimes C + B \otimes C$. In the following sections, we will mostly deal with square diagonal matrices, therefore, it is convenient to use the following notation. For diagonal matrices $S (2\times 2)$ and $T (m\times m)$, only taking into account the diagonal entries, one can write $S = diag(s_{11}, s_{22})$ and $T = diag(t_{11},... ,t_{mm})$, with Kronecker product as,
\begin{equation}
\begin{split}
S \otimes T &= diag(s_{11}t_{11},.. .,s_{11}t_{mm},s_{22}t_{11},...,s_{22}t_{mm}).
\end{split}
\end{equation}
Consider two square matrix $S$ and T  of order $m$ and $n$ respectively, Kronecker sum ($\oplus$) operation is defined as \cite{britanak2010discrete},
\begin{equation}
    S \oplus T = S \otimes I_{m} + I_{n} \otimes T, 
\end{equation}
where, $I_{x}$ is an Identity matrix of order $x$. As evident, the dimension of $S \oplus T$ is $mn$, therefore similar to the Kronecker product, Kronecker sum also increases the dimension. It is also non-commutative and associative operator, however under the product Kronecker sum is not distributive over the usual addition i.e., $(A + B) \oplus C \neq A\oplus C + B \oplus C$. For diagonal matrices  $S = diag(s_{11}, s_{22})$ and $T = diag(t_{11},... ,t_{mm})$, it can be compactly written as
\begin{equation}
\begin{split}
S \oplus T = & diag((s_{11} + t_{11}), . . ,(s_{11} + t_{mm}),\\
            &\quad \quad \quad\;\;(s_{22} + t_{11}),...,(s_{22}+t_{mm})).
\end{split}
\end{equation}
In the following sections, we make use of these two operations to represent Hamiltonian of spin chain systems in various scenarios, leading to an efficient and simple procedure to obtain partition function without explicit consideration of the involved microstates.

\section{Representing Spin interactions}
In this work, we restrict our discussion to spin-half particles. There are two eigenstates corresponding to spin-half particles, spin-up ($\uparrow$) and spin-down ($\downarrow$).Therefore, microstates of a spin-half particle is given by the set $\{\uparrow,\downarrow\}$. Matrix representation of the microstates of a spin-half particle is given by,
\begin{equation}
\label{mat}
S = 
\begin{bmatrix}
+1& 0\\
0& -1\\
\end{bmatrix},
\end{equation}
where the values +1 and -1 corresponds to spin-up and spin-down states, respectively 
In this section we use $(\uparrow,\downarrow)$ to represent the states which illustrate the counting of microstates using the Kronecker sum and Kronecker product for a given interaction. In the following, we consider the Hamiltonian involving product of spins, sum of spins and finally a combination of both.
\newline\\
\textbf{Case 1.\quad Product of spins}\\
To begin with,  consider a system of two spins with the Hamiltonian given by,
\begin{equation}
    H = S_{1}S_{2}.
\end{equation}
Since each spin can independently be in Spin-up and Spin-down state,  microstates of this interaction is given by the set $\{\uparrow \uparrow, \uparrow \downarrow, \downarrow \uparrow, \downarrow \downarrow\}$.
The partition function is obtained as,
\begin{equation}
\begin{split}
Z &= \sum_{microstates} \exp{(-\beta H)} = \sum_{S_{1},S_{2}} \exp{(-\beta S_{1}S_{2})}\\
  &= e^{-\beta \uparrow \uparrow} + e^{-\beta \uparrow \downarrow}+e^{-\beta \downarrow \uparrow} + e^{-\beta \downarrow \downarrow}.
\end{split}
\end{equation}
It is to be noted that in the above we have explicitly considered all the microstates involving configuration of individual spins.  The above procedure can be modelled by using a Kronecker product operation.
We can write the matrix representation of both spins as, $S_{1} = diag(\uparrow, \downarrow)$ and $S_{2} = diag(\uparrow, \downarrow)$ and observing that the Kronecker product between them yields $ S_{1} \otimes S_{2} = diag(\uparrow \uparrow, \uparrow \downarrow, \downarrow \uparrow, \downarrow \downarrow)$.
The partition function is then given by,
\begin{equation}
\begin{split}
    Z &= \operatorname{Tr}(\exp(-\beta S_{1} \otimes S_{2}))\\
      &= e^{-\beta \uparrow \uparrow} + e^{-\beta \uparrow \downarrow}+e^{-\beta \downarrow \uparrow} + e^{-\beta \downarrow \downarrow}.
\end{split}
\end{equation}
Therefore, one can represent all the microstates of the Hamiltonian, $H = S_{1}S_{2}$ as a diagonal matrix $\hat{H} = S_{1} \otimes S_{2}$.
In general, the interaction of n-spins of type $\mathcal{H} = S_{1}S_{2}S_{3}...S_{n}$ having $2^{n}$ microstates can be represented by,
\begin{equation}
    \hat{\mathcal{H}} = S_{1} \otimes S_{2} \otimes S_{3}\otimes....\otimes S_{n}.
\end{equation}
The partition function is given by,
\begin{equation*}
    Z = \operatorname{Tr}(\operatorname{e}^{-\beta \hat{\mathcal{H}}}).
\end{equation*}
Therefore, the Kronecker product operation represents the microstates of the product of the independent spins.
\newline
\\
\textbf{Case 2.\quad Sum of spins}\\
Consider a two spin system with the Hamiltonian given by, $H = S_{1} + S_{2}$.
Through explicit counting one obtains the microstates of this interaction as $ \{(\uparrow + \uparrow), (\uparrow + \downarrow), (\downarrow + \uparrow), (\downarrow + \downarrow)\}$. Interestingly, these microstates can be modelled using Kronecker sum as,
\begin{equation}
\begin{split}
        \hat{H} &= S_{1} \oplus S_{2}\\
                &= diag(\uparrow, \downarrow) \oplus diag(\uparrow, \downarrow)\\
                &= diag((\uparrow + \uparrow), (\uparrow + \downarrow), (\downarrow + \uparrow), (\downarrow + \downarrow)).
\end{split}
\end{equation}
We can generalise the above for $n-spins$  with the Hamiltonian  $H = S_{1}+S_{2}+...+S_{n}$ having $2^{n}$ microstates  through,
\begin{equation}
    \hat{H} = S_{1} \oplus S_{2} \oplus...\oplus S_{n}.
\end{equation}
Therefore, Kronecker sum $(\oplus)$ operation represents the microstates of the sum of the independent spins.
\newline
\\
\textbf{Case 3.\quad Sum of product of spins}\\
For a more general case where we have both product and sum in a Hamiltonian, for example, interaction of type $H = S_{1}S_{2} + S_{3}$ for a three spin system which have the $2^{3} = 8$ microstates, through explicit counting one obtain the microstates as following,
\begin{equation}
\begin{split}
    \{&(\uparrow \uparrow + \uparrow), (\uparrow \uparrow + \downarrow), (\uparrow \downarrow + \uparrow), (\uparrow \downarrow + \downarrow),\\ &(\downarrow \uparrow + \uparrow), (\downarrow \uparrow + \downarrow), (\downarrow \downarrow + \uparrow), (\downarrow \downarrow + \downarrow)\}.
\end{split}
\end{equation}
It is now straightforward to obtain the above through Kronecker sum and product by simply using previous cases one after the other.
First, there is a product between $S_{1}$ and $S_{2}$ followed by a Kronecker sum with $S_{3}$.
The matrix representing the microstates is then given by,
\begin{equation*}
\begin{split}
    \hat{H} &= (S_{1}\otimes S_{2})\oplus S_{3}\\
            &= diag(\uparrow \uparrow, \uparrow \downarrow, \downarrow \uparrow, \downarrow \downarrow) \oplus diag(\uparrow, \downarrow)\\
            &= diag((\uparrow \uparrow + \uparrow), (\uparrow \uparrow + \downarrow), (\uparrow \downarrow + \uparrow), (\uparrow \downarrow + \downarrow),\\ 
            &\quad \quad \quad \;\;(\downarrow \uparrow + \uparrow), (\downarrow \uparrow + \downarrow), (\downarrow \downarrow + \uparrow), (\downarrow \downarrow + \downarrow))
\end{split}
\end{equation*}
In the above cases, we are explicitly counting the microstates but in an organized way.
Thus, we can represent the microstates of an interaction of independent spins by replacing product with tensor product and sum with Kronecker sum. The resultant diagonal matrix gives all microstates corresponding to the interaction under consideration.

\section{Spins in the presence of magnetic field}
Here, we make use of the approach described in previous section to obtain explicit expressions for partition functions for system of mutually non interacting spins present in an external magnetic field. The Hamiltonian for a spin $S_{1}$ in the external magnetic field is given by,
\begin{equation}
    H = -kS_{1},
\end{equation}
where, k is a positive constant \cite{baxter2007exactly}.
\newline
For the spin-half case, $S_{1}$ can take two configurations $\{\uparrow, \downarrow\}$. Adding one more spin $S_{2}$ in this system and not considering the mutual interaction between spins, the Hamiltonian for this system  in the presence of  external magnetic field is given by,
\begin{equation}
    H = -k(S_{1} + S_{2}).
\end{equation}
As discussed in the previous section and using matrix representation of Eq. \ref{mat}, the microstates of this system are obtained through,
\begin{equation}
\begin{split}
    \hat{H} &= -k(S_{1} \oplus S_{2})\\
            &= -k\; diag(1, -1) \oplus diag(1,-1)\\
            &= -k\;diag(2,0,0,-2).
\end{split}
\end{equation}
It is now straightforward to evaluate the partition function. One obtains,
\begin{equation*}
\begin{split}
        Z &= \operatorname{Tr}(\exp(-\beta \hat{H})\\
          &= \operatorname{Tr}(\exp(\beta\;diag(2k,0,0,-2k))).
\end{split}
\end{equation*}
From the above diagonal matrix, one observes that the degeneracy of each energy level are as follows: energy level $2k$,  $0$, $-2k$   have the degeneracy 1, 2, and 1, respectively.  Therefore, the expression for the partition function evaluates to,
\begin{equation}
\begin{split}
        Z &=  2 + e^{2k\beta} + e^{-2k\beta}\\
          &= 2(1 +\cosh{(2k\beta)}).
\end{split}
\end{equation}
Further adding one more spin into the system, the partition function becomes,
\begin{equation*}
\begin{split}
        Z &= \operatorname{Tr}(\exp(k\beta \;diag(2,0,0,-2)\oplus diag(1,-1))\\
          &= \operatorname{Tr}(\exp(k\beta \;diag(3,1,1,-1,1,-1,-1,-3)))\\
\end{split}
\end{equation*}
where degeneracy corresponding to the energies $3k$, $1k$, $-1k$, and $-3k$ are given by 1, 3, 3 and 1 respectively. Therefore, we observe that using this Kronecker sum and product structure simplifies the counting of degeneracy corresponding to various energy levels. Since it does not explicitly refer to the specific configuration, counting of degeneracy reduces to identifying equivalent elements in a diagonal matrix. With this, the partition function evaluates to
\begin{equation}
\begin{aligned}
    Z &= e^{3k\beta}+3e^{1k\beta}+3e^{-1k\beta}+e^{-3k\beta}\\
    &= 2(\cosh{3k\beta} + 3\cosh{k\beta})
\end{aligned}    
\end{equation}
For $n$ mutually non-interacting spins in an external magnetic field, the Hamiltonian is written as,
\begin{equation}
    \hat{H} = k(S_{1}\oplus S_{2}\oplus S_{3}\oplus...\oplus S_{n})
\end{equation}

\section{1-d Ising model}
In the Ising interaction, we consider spin-lattice and each spin interacts only with its neighboring spin by a product type interaction. Hamiltonian of the Ising model is given by $H = -j\;\sum _{\langle ik \rangle}S_{i}S_{k}$, where $\langle ik \rangle$ represents sum over neighboring spins and $j$ is a  coupling constant \cite{baxter2007exactly}. In the following, we consider various configurations and provide a prescription to find the partition function employing Kronecker sum and product.
\newline\\
\textbf{Case 1.  1-d open-chain}

Consider a three spins chain as shown in the Fig. 1. The Hamiltonian for this system is given by
\begin{equation}\label{threespins}
    H = -j(S_{1}S_{2} + S_{2}S_{3}). 
\end{equation}
Here, the spin $S_{2}$ is a common term between $S_{1}$ and $S_{3}$. It is not immediately obvious how to write this interaction in terms of Kronecker product and Kronecker sum.
One observes the microstates of this system as
\begin{equation}\label{micro1}
\begin{split}
    [&(\uparrow \uparrow + \uparrow \uparrow),(\uparrow \uparrow + \uparrow \downarrow),(\uparrow \downarrow + \downarrow \uparrow),(\uparrow \downarrow + \downarrow \downarrow),\\
    &(\downarrow \uparrow + \uparrow \uparrow),(\downarrow \uparrow + \uparrow \downarrow),(\downarrow \downarrow + \downarrow \uparrow),(\downarrow \downarrow + \downarrow \downarrow)]
\end{split}
\end{equation}

\begin{figure}[htbp]
    \centering
    \includegraphics[scale = 0.4]{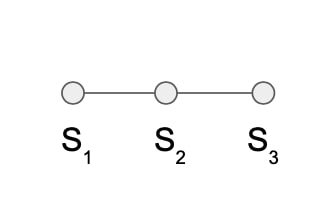}
    \caption{Ising Interaction 3 spins}
    \label{FIG.1}
\end{figure}

It is evident that for the Hamiltonian given by (\ref{threespins}), the two terms,  $S_{1}S_{2}$ and $S_{2}S_{3}$ can only take values +1 and -1. Moreover, it is the value of $S_{3}$ which determines the product term $S_{2}S_{3}$, for specific choices of the values of $S_{1}$ and $S_{2}$. Therefore, the microstates in (\ref{micro1}) is equivalent to the following,
\begin{equation}
\begin{split}
    [&(\uparrow \uparrow + \uparrow),(\uparrow \uparrow + \downarrow),(\uparrow \downarrow + \downarrow),(\uparrow \downarrow + \uparrow),\\
    &(\downarrow \uparrow + \uparrow),(\downarrow \uparrow + \downarrow),(\downarrow \downarrow + \downarrow),(\downarrow \downarrow + \uparrow)]
\end{split}
\end{equation}
Therefore, value of $S_{3}$ can be independently added to the first term $S_{1}S_{2}$.
In terms of the values corresponding to the microstates, the Hamiltonian $H = -j(S_{1}S_{2} + S_{2}S_{3})$ are same as of
and $H = -j(S_{1}S_{2} + S_{3})$ are equivalent. In term of Kronecker sum and Kronecker product, the Hamiltonian (\ref{threespins}) can be written as,
\begin{equation}
    \hat{H} = -j((S_{1}\otimes S_{2})\oplus S_{3})
\end{equation}
If we add one more spin to this linear chain, we get the following Hamiltonian,
\begin{equation}
    H = -j(S_{1}S_{2} +S_{2}S_{3} + S_{3}S_{4} )
\end{equation}
\begin{figure}[htbp]
    \centering
    \includegraphics[scale = 0.3]{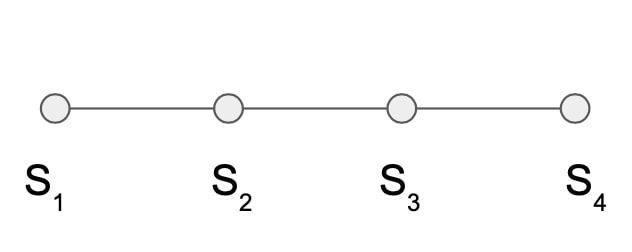}
    \caption{Ising spin interaction of 4 spins}
    \label{FIG.2}
\end{figure}
Again we can do the same thing, this time we first fix $S_{1}$ and $S_{2}$, we can see that the second term only depends on the value of $S_{3}$, and then fixing that value of $S_{3}$, we permute over $S_{4}$. It is represented by,
\begin{equation}
    H = -j(S_{1}S_{2} + S_{3} + S_{4})
\end{equation}
In term of Kronecker sum and Kronecker product, the above Hamiltonian can be written as,
\begin{equation}
    \hat{H} = -j((S_{1}\otimes S_{2})\oplus S_{3} \oplus S_{4})
\end{equation}

Carrying on in the same way, we can write the $1-d$ linear open-chain Ising model of $n-$spins in Kronecker product and Kronecker sum as:
\begin{equation}
    \hat{H} = -j\{(S_{1}\otimes S_{2})\oplus S_{3}\oplus....\oplus S_{n}\}
\newline
\end{equation}
\textbf{Case 2.  1-d closed-chain}

For a $1-d$ closed-chain Ising model of three spins, the Hamiltonian is given by
\begin{equation}
    H = -j(S_{1}S_{2} + S_{2}S_{3} + S_{3}S_{1}).
\end{equation}
\begin{figure}[htbp]
    \centering
    \includegraphics[scale = 0.3]{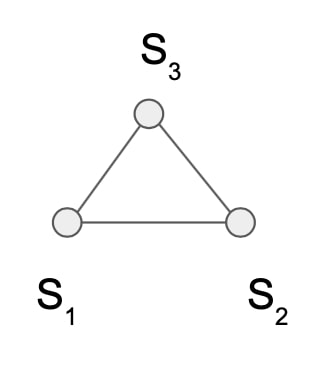}
    \caption{closed-chain Ising interaction of 3 spins}
    \label{FIG.3}
\end{figure}
There are three independent spins: $S_{1}$, $S_{2}$ and $S_{3}$. Therefore, the total number of microstates are, $2^{3} = 8$.
The microstates of this system are:
\begin{equation*}
    [3, -1, -1, -1, -1, -1, -1, 3]
\end{equation*}
As shown earlier, for three spins as open-chain, one can represent the microstates by,
\begin{equation}
    \hat{H} = -j((S_{1}\otimes S_{2})\oplus S_{3})
\end{equation}
 To describe three spins closed-chain system, we cannot directly use Kronecker sum and Kronecker product, like 
\begin{equation*}
    \hat{H} = -j((S_{1}\otimes S_{2})\oplus S_{3} \oplus S_{1}),
\end{equation*}
or
\begin{equation*}
    \hat{H} = -j((S_{1}\otimes S_{2})\oplus S_{3} \otimes S_{1}).
\end{equation*}
It is because these expressions will yield the number of microstates as $2^{4}$. It will not describe a $1-d$ closed-loop problem of 3 spins, which require $8$ microstates.

The only way we can proceed from an open-chain to closed-chain is by adding an $8 \times 8$ diagonal matrix ($D_{3}$) to the open-chain expression as following,
\begin{equation}
    \hat{H} = -j((S_{1}\otimes S_{2})\oplus S_{3} \;+\; D_{3} ).
\end{equation}
In a similar manner, for the case of four spins, the Hamiltonian is given by
\begin{equation}
    H = -j(S_{1}S_{2} + S_{2}S_{3} + S_{3}S_{4} + S_{4}S_{1}).
\end{equation}
With 4 independent spins, the total number of microstates are $2^{4} = 16$.
To describe the closed-chain of $4$ spins, we have to add $16 \times 16$ diagonal matrix ($D_{4}$) to the open-chain expression as,
\begin{equation}
    \hat{H} = -j((S_{1}\otimes S_{2})\oplus S_{3} \oplus S_{4} \;+ \;D_{4} ).
\end{equation}
\begin{figure}[htbp]
    \centering
    \includegraphics[scale = 0.3]{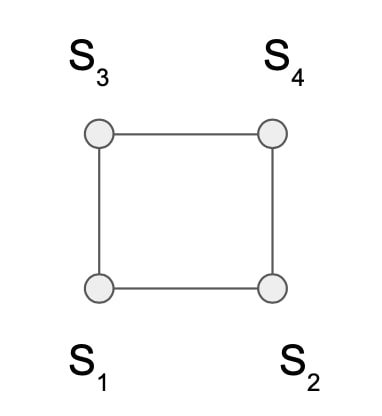}
    \caption{closed-chain Ising interaction of 4 spins}
    \label{FIG.4}
\end{figure}
And for $n$ spins closed-chain, we have to add an $2^{n} \times 2^{n}$ diagonal matrix $(D_{n})$ to the open-chain term,
\begin{equation}
        \hat{H} = -j((S_{1}\otimes S_{2})\oplus S_{3}\oplus....\oplus S_{n} \;+\; D_{n}).
\end{equation}
It turns out that the only consistent way of generalizing the expression of closed-chain from open-chain, is to choose 
\begin{equation}
    D_{n} = S_{1} \otimes S_{2} \otimes S_{3}\otimes....\otimes S_{n}.
\end{equation}
Therefore, for closed-chain 3 spins, the microstates are represented by the expression:
\begin{equation}
\begin{split}
    \hat{H} &= -j\{(S_{1}\otimes S_{2})\oplus S_{3}\;+\; D_{3} \}\\
      &= -j\{(S_{1}\otimes S_{2})\oplus S_{3}\;+\;S_{1} \otimes S_{2} \otimes S_{3} \}\\ 
      &= diag(3, -1, -1, -1, -1, -1, 3, -1).
\end{split}
\end{equation}
To describe the closed-chain Ising model of n-spins, one has to add a diagonal matrix $(D_{n})$ to the open-chain expression. The correct diagonal matrix is the one which generalizes closed-chain expression from open-chain for n-spins. It turns out that there is only one such expression given by,
\begin{equation*}
    D_{n} = S_{1} \otimes S_{2} \otimes S_{3}\otimes....\otimes S_{n}.
\end{equation*}
\newline
\textbf{Case 3. Open-chain in the presence of magnetic field}

Ising model in the presence of external magnetic field has the following expression \cite{baxter2007exactly}-
Considering 3 spins, the Hamiltonian is given by,
\begin{equation}
\begin{split}
    H &= Ising\;term\; +\; Magnetic\;field\;term\\
      &= -j(S_{1}S_{2} + S_{2}S_{3}) -k(S_{1} + S_{2} + S_{3}).
\end{split}
\end{equation}
The microstates of this system are,
\begin{equation}
\begin{split}
    [&(-2j-3k),(-k),(2j-k),(+k),\\
     &(-k),(2j+k),(+k),(-2j+3k)]
\end{split}
\end{equation}
Yet we have dealt with Ising term and magnetic field term separately and obtained the corresponding Hamiltonians in terms of Kronecker sum and Kronecker product.
To write open-chain Ising model in presence of external magnetic field, the obvious way to proceed is to use previous methods directly:
\begin{equation*}
\begin{split}
    \hat{H} &= -j((S_{1} \otimes S_{2})\oplus S_{3}) - k(S_{1}\oplus S_{2} \oplus S_{3})\\
      &= -j\;diag(2,0,0,-2,0,-2,2,0)\\ & \quad -k\;diag(3,1,1,-1,1,-1,-1,-3)
\end{split}
\end{equation*}
Here, we can see that the Ising term and the magnetic term individually give the correct energies and degeneracy: $diag(2,0,0,-2,0,-2,2,0)$ gives the correct count for microstates of $S_{1}S_{2} + S_{2}S_{3}
$ and $diag(3,1,1,-1,1,-1,-1,-3)$ gives the correct count for microstates of $S_{1} + S_{2} + S_{3}$. 

But, after adding them we get the microstates :
\begin{equation}
\begin{split}
    \hat{H} = diag(&(-2j-3k),(-k),(-k),(2j+k),\\ & (-k),(2j+k),(-2j+k),(3k))
\end{split}
\end{equation}
Clearly, these are not the microstates corresponding to the system-
\begin{equation}
    H = -j(S_{1}S_{2} + S_{2}S_{3}) -k(S_{1} + S_{2} + S_{3})
\end{equation}
Thus, we have correct count of microstates in the Ising term and in magnetic term. But the order of the diagonal elements doesn't match to give the correct microstates.\\
To resolve this problem we have to identify the correct element order of Ising term that matches up with magnetic term and perform a similarity transformation which can be generalized to n-spins. 

The correct order of the diagonal elements in this case is:
\begin{equation}
    diag(2,0,-2,0,0,-2,0,2)
\end{equation}
Let $M_{3}$ be the transformation matrix such that,
\begin{equation*}
\begin{split}
    &diag(2,0,-2,0,0,-2,0,2) = M^{-1}_{3}((S_{1}\otimes S_{2})\oplus S_{3})M_{3}\\
    &diag(2,0,-2,0,0,-2,0,2) = M^{-1}_{3}diag(2,0,0,-2,\\
    &\quad \quad \quad \quad \quad \quad \quad \quad \quad \quad \quad \quad \quad \quad \quad \quad \quad \; 0,-2,2,0)M_{3}
\end{split}
\end{equation*}
It turns out that the matrix $M_{3}$ is block diagonal, with each block a $2 \times 2$ matrix.
\begin{equation}
    M_{3} = diag(I_{2},\sigma_{x},I_{2},\sigma_{x})
\end{equation}
where $I_{2}$ = 
$\begin{bmatrix}
1 & 0 \\
0 & 1
\end{bmatrix}$
and 
$\sigma_{x}$ = 
$\begin{bmatrix}
0 & 1\\
1 & 0
\end{bmatrix}$
\newline
\\
It is straightforward to check that $M_{3}^{2} = I_{8}$. So $M^{-1}_{3} = M_{3}$, where $I_{8}$ is identity matrix of order $8$.
\newline\\
So the correct representation of microstates is given by:
\begin{equation*}
\begin{split}
    \hat{H} &= -j\{M_{3} \left[(S_{1}\otimes S_{2})\oplus S_{3}\right]M_{3}\} -k(S_{1}\oplus S_{2}\oplus S_{3})\\
\end{split}
\end{equation*}
We will call the modified Ising term as $G_{3}$:
\begin{equation}
    G_{3} = M_{3}\left[(S_{1}\otimes S_{2})\oplus S_{3}\right]M_{3}
\end{equation}
This evaluates to,
\begin{equation}
\begin{split}
    \hat{H} &= -jG_{3} -k(S_{1}\oplus S_{2}\oplus S_{3})\\
    &= -j\; diag(2,0,-2,0,0,-2,0,2)\\ 
      &\quad- k\; diag(3,1,1,-1,1,-1,-1,-3)\\
      &= diag((-2j-3k),(-k),(2j-k),(+k),\\
      &\quad \quad \quad \;\; \; (-k),(2j+k),(+k),(-2j+3k))
\end{split}
\end{equation}
If we add one more spin in the system, then  the Hamiltonian can be written in terms of Kronecker sum and Kronecker product by repeated use of similarity transformation  :
\begin{equation}
\begin{split}
    \hat{H} &= -j\{M_{4}(G_{3}\oplus S_{4})M_{4}\}\\
      & \quad -k(S_{1}\oplus S_{2} \oplus S_{3} \oplus S_{4})
\end{split}
\end{equation}
where $M_{4} = diag(I_{2},\sigma_{x},I_{2},\sigma_{x},I_{2},\sigma_{x},I_{2},\sigma_{x})$

We can call this modified Ising term $G_{4}$:
\begin{equation}
    G_{4} = M_{4}(G_{3}\oplus S_{4})M_{4}
\end{equation}
We have to repeatedly use similarity transform on the Ising term to get the correct microstates for an Ising interaction in external magnetic field.
This can be generalised to a system containing $n$-spins.
\begin{equation}
\begin{split}
    \hat{H} &= -jG_{n} - k(S_{1}\oplus S_{2} \oplus S_{3}\oplus...\oplus S_{n})
\end{split}
\end{equation}
where, $G_{n} = M_{n}(G_{n-1}\oplus S_{n})M_{n}$.\\
The transformation matrix $M_{n}$ only contains sequence of repeated block matrices: $I_{2}$ and $\sigma_{x}$, we can write this in a compact form:
\begin{equation*}
    M_{n} = diag((I_{2},\sigma_{x})_{2^{n-2}})\quad and\quad
    M_{2} = I_{4}
\end{equation*}
where $(I_{2},\sigma_{x})_{2^{n-2}}$ means that the sequence $(I_{2},\sigma_{x})$ is repeated $2^{n-2}$ times and $I_{4}$ is an identity matrix of order 4.
E.g. 
\begin{equation*}
\begin{split}
        M_{2} &= diag(1,1,1,1)\\
        M_{3} &= diag(I_{2},\sigma_{x},I_{2},\sigma_{x})\\
        M_{4} &= diag(I_{2},\sigma_{x},I_{2},\sigma_{x},I_{2},\sigma_{x},I_{2},\sigma_{x})
\end{split}
\end{equation*}
\textbf{Case 4. closed-chain in the presence of magnetic field}

To represent the closed-chain Ising model in magnetic field, we follow the same procedure we followed to go from 1-d open-chain to 1-d closed-chain in the absence of magnetic field. We add a $2^{n} \times 2^{n}$ diagonal matrix $(D_{n})$ to the Ising term.
\begin{equation}
    \hat{H} = -j(G_{n} + D_{n}) - k(S_{1}\oplus S_{2} \oplus S_{3}\oplus...\oplus S_{n})
\end{equation}
Here again it turns out that the only consistent way of generalizing the expression to $n$- spins for closed-chain in external magnetic field is to choose 
\begin{equation*}
    D_{n} = {S_{1}\otimes I_{2^{n-2}} \otimes S_{n}}
\end{equation*}
Where $I_{2^{n-2}}$ is an identity matrix of order $2^{n-2}$.\\
E.g. for $3$ spins, the microstates are given by the diagonal entries of
\begin{equation}
\begin{split}
    \hat{H} &= -j(G_{3} + D_{3}) - k(S_{1}\oplus S_{2} \oplus S_{3})\\
      &= -j(G_{3} \;+\; S_{1}\otimes I_{2} \otimes S_{3}) - k(S_{1}\oplus S_{2} \oplus S_{3})
\end{split}
\end{equation}
E.g. for $4$ spins, the microstates are given by the diagonal entries of 
\begin{equation}
\begin{split}
    \hat{H} &= -j(G_{4} + D_{4}) - k(S_{1}\oplus S_{2} \oplus S_{3}\oplus S_{4})\\
      &= -j(G_{4} \;+\; S_{1}\otimes I_{4} \otimes S_{4})\\
      &\quad - k(S_{1}\oplus S_{2} \oplus S_{3} \oplus S_{4})
\end{split}
\end{equation}

\section{Conclusions}
In this paper we have provided a method of representing microstates of various spin interactions in the 1-d Ising model using Kronecker sum $(\oplus)$ and Kronecker product $(\otimes)$ operations on matrices. The partition function, which gives all the relevant information about the system being studied, is found out by taking the trace of the exponential of the resultant matrix.

This method was applied to open and closed 1-d chains and we readily obtained the correct values. We also applied this approach for spin interactions with external magnetic field and were able to find the correct count for the Ising and magnetic term separately. Using a similarity transformation we corrected for the order of terms in the diagonal matrix and obtained results which match with the literature.

Solutions in 1-d Ising models involve counting of microstates which can get difficult to keep track if not done systematically. Our system provides a systematic way of doing such calculations while also giving an intuitive grasp of the underlying mechanism and we believe that this method can find useful applications in undergraduate classrooms as well as practicing researchers in the field because of its computational friendly formalism. 

\section*{Acknowledgement}

We wish to thank Dr. Pradeep Kumar Mohanty, Dr.
Bhavtosh Bansal and Dr. Rumi De from IISER Kolkata for their fruitful discussions. We also wish to thank Department of Science
and Technology, Government of India for providing financial support as INSPIRE fellowship.
%\vskip 2 cm
\section*{References}
\bibliographystyle{samp}

\bibliography{sample}

\begin{thebibliography}{10}
\providecommand{\url}[1]{{#1}}
\providecommand{\urlprefix}{URL }
\expandafter\ifx\csname urlstyle\endcsname\relax
  \providecommand{\doi}[1]{DOI~\discretionary{}{}{}#1}\else
  \providecommand{\doi}{DOI~\discretionary{}{}{}\begingroup
  \urlstyle{rm}\Url}\fi

\bibitem{RevModPhys.39.883}
BRUSH, S.G.: History of the lenz-ising model.
\newblock Rev. Mod. Phys. \textbf{39}, 883--893 (1967).
\newblock \doi{10.1103/RevModPhys.39.883}.
\newblock \urlprefix\url{https://link.aps.org/doi/10.1103/RevModPhys.39.883}

\bibitem{1925ZPhy...31..253I}
{Ising}, E.: {Beitrag zur Theorie des Ferromagnetismus}.
\newblock Zeitschrift fur Physik \textbf{31}(1), 253--258 (1925).
\newblock \doi{10.1007/BF02980577}

\bibitem{10.2307/24097007}
Bhattacharjee, S.M., Khare, A.: Fifty years of the exact solution of the
  two-dimensional ising model by onsager.
\newblock Current Science \textbf{69}(10), 816--821 (1995).
\newblock \urlprefix\url{http://www.jstor.org/stable/24097007}

\bibitem{PhysRev.60.252}
Kramers, H.A., Wannier, G.H.: Statistics of the two-dimensional ferromagnet.
  part i.
\newblock Phys. Rev. \textbf{60}, 252--262 (1941).
\newblock \doi{10.1103/PhysRev.60.252}.
\newblock \urlprefix\url{https://link.aps.org/doi/10.1103/PhysRev.60.252}

\bibitem{PhysRev.65.117}
Onsager, L.: Crystal statistics. i. a two-dimensional model with an
  order-disorder transition.
\newblock Phys. Rev. \textbf{65}, 117--149 (1944).
\newblock \doi{10.1103/PhysRev.65.117}.
\newblock \urlprefix\url{https://link.aps.org/doi/10.1103/PhysRev.65.117}

\bibitem{10.2307/96591}
Williams, E.J.: The effect of thermal agitation on atomic arrangement in
  alloys. iii.
\newblock Proceedings of the Royal Society of London. Series A, Mathematical
  and Physical Sciences \textbf{152}(875), 231--252 (1935).
\newblock \urlprefix\url{http://www.jstor.org/stable/96591}

\bibitem{1935RSPSA.150..552B}
{Bethe}, H.A.: {Statistical Theory of Superlattices}.
\newblock Proceedings of the Royal Society of London Series A
  \textbf{150}(871), 552--575 (1935).
\newblock \doi{10.1098/rspa.1935.0122}

\bibitem{Kawamura_2010}
Kawamura, H.: Two models of spin glasses {\textemdash} ising versus heisenberg.
\newblock Journal of Physics: Conference Series \textbf{233}, 012012 (2010).
\newblock \doi{10.1088/1742-6596/233/1/012012}.
\newblock \urlprefix\url{https://doi.org/10.1088/1742-6596/233/1/012012}

\bibitem{Hopfield2554}
Hopfield, J.J.: Neural networks and physical systems with emergent collective
  computational abilities.
\newblock Proceedings of the National Academy of Sciences \textbf{79}(8),
  2554--2558 (1982).
\newblock \doi{10.1073/pnas.79.8.2554}.
\newblock \urlprefix\url{https://www.pnas.org/content/79/8/2554}

\bibitem{Schneidman06weakpairwise}
Schneidman, E., Ii, M.J.B., Segev, R., Bialek, W.: Weak pairwise correlations
  imply strongly correlated network states in a neural population. (2006)

\bibitem{2015JPhCS.574a2149T}
{Takaishi}, T.: {Multiple Time Series Ising Model for Financial Market
  Simulations}.
\newblock In: Journal of Physics Conference Series, \emph{Journal of Physics
  Conference Series}, vol. 574, p. 012149 (2015).
\newblock \doi{10.1088/1742-6596/574/1/012149}

\bibitem{doi:10.1119/1.2779882}
Stauffer, D.: Social applications of two-dimensional ising models.
\newblock American Journal of Physics \textbf{76}(4), 470--473 (2008).
\newblock \doi{10.1119/1.2779882}.
\newblock \urlprefix\url{https://doi.org/10.1119/1.2779882}

\bibitem{doi:10.1142/S0129183101001845}
BORNHOLDT, S.: Expectation bubbles in a spin model of markets: Intermittency
  from frustration across scales.
\newblock International Journal of Modern Physics C \textbf{12}(05), 667--674
  (2001).
\newblock \doi{10.1142/S0129183101001845}.
\newblock \urlprefix\url{https://doi.org/10.1142/S0129183101001845}

\bibitem{2021PhyA..57325963M}
{Mello}, I.F., {Squillante}, L., {Gomes}, G.O., {Seridonio}, A.C., {de Souza},
  M.: {Epidemics, the Ising-model and percolation theory: A comprehensive
  review focused on Covid-19}.
\newblock Physica A Statistical Mechanics and its Applications \textbf{573},
  125963 (2021).
\newblock \doi{10.1016/j.physa.2021.125963}

\bibitem{2004PhRvA..70e2322L}
Lee, C.F., Johnson, N.F.: Efficient quantum computation within a disordered
  heisenberg spin chain.
\newblock Phys. Rev. A \textbf{70}, 052322 (2004).
\newblock \doi{10.1103/PhysRevA.70.052322}.
\newblock \urlprefix\url{https://link.aps.org/doi/10.1103/PhysRevA.70.052322}

\bibitem{2016NatCo...713070M}
{Marchukov}, O.V., {Volosniev}, A.G., {Valiente}, M., {Petrosyan}, D.,
  {Zinner}, N.T.: {Quantum spin transistor with a Heisenberg spin chain}.
\newblock Nature Communications \textbf{7}, 13070 (2016).
\newblock \doi{10.1038/ncomms13070}

\bibitem{PhysRevA.70.032313}
\ifmmode \check{S}\else \v{S}\fi{}telmachovi\ifmmode~\check{c}\else \v{c}\fi{},
  P., Bu\ifmmode~\check{z}\else \v{z}\fi{}ek, V.: Quantum-information approach
  to the ising model: Entanglement in chains of qubits.
\newblock Phys. Rev. A \textbf{70}, 032313 (2004).
\newblock \doi{10.1103/PhysRevA.70.032313}.
\newblock \urlprefix\url{https://link.aps.org/doi/10.1103/PhysRevA.70.032313}

\bibitem{baxter2007exactly}
Baxter, R.: Exactly Solved Models in Statistical Mechanics.
\newblock Dover books on physics. Dover Publications (2007).
\newblock \urlprefix\url{https://books.google.co.in/books?id=G3owDULfBuEC}

\bibitem{PhysRev.88.1332}
Kac, M., Ward, J.C.: A combinatorial solution of the two-dimensional ising
  model.
\newblock Phys. Rev. \textbf{88}, 1332--1337 (1952).
\newblock \doi{10.1103/PhysRev.88.1332}.
\newblock \urlprefix\url{https://link.aps.org/doi/10.1103/PhysRev.88.1332}

\bibitem{costa2003f}
Costa, G., Maciel, A.L.: Combinatorial formulation of ising model revisited
  \textbf{25}(1) (2003).
\newblock \doi{10.1590/S1806-11172003000100007}.
\newblock \urlprefix\url{https://doi.org/10.1590/S1806-11172003000100007}

\bibitem{PARK2022107325}
Park, J., Jin, I.H., Schweinberger, M.: Bayesian model selection for
  high-dimensional ising models, with applications to educational data.
\newblock Computational Statistics \& Data Analysis \textbf{165}, 107325
  (2022).
\newblock \doi{https://doi.org/10.1016/j.csda.2021.107325}.
\newblock
  \urlprefix\url{https://www.sciencedirect.com/science/article/pii/S0167947321001596}

\bibitem{1993JSP}
{Gofman}, M., {Adler}, J., {Aharony}, A., {Harris}, A.B., {Stauffer}, D.:
  {Series and Monte Carlo study of high-dimensional Ising models}.
\newblock Journal of Statistical Physics \textbf{71}(5-6), 1221--1230 (1993).
\newblock \doi{10.1007/BF01049970}

\bibitem{10.1214/09-AOS691}
Ravikumar, P., Wainwright, M.J., Lafferty, J.D.: {High-dimensional Ising model
  selection using l1-regularized logistic regression}.
\newblock The Annals of Statistics \textbf{38}(3), 1287 -- 1319 (2010).
\newblock \doi{10.1214/09-AOS691}.
\newblock \urlprefix\url{https://doi.org/10.1214/09-AOS691}

\bibitem{PhysRevLett.62.361}
Wolff, U.: Collective monte carlo updating for spin systems.
\newblock Phys. Rev. Lett. \textbf{62}, 361--364 (1989).
\newblock \doi{10.1103/PhysRevLett.62.361}.
\newblock \urlprefix\url{https://link.aps.org/doi/10.1103/PhysRevLett.62.361}

\bibitem{2016}
Ibarra-Garc{\'{\i}}a-Padilla, E., Malanche-Flores, C.G., Poveda-Cuevas, F.J.:
  The hobbyhorse of magnetic systems: the ising model \textbf{37}(6), 065103
  (2016).
\newblock \doi{10.1088/0143-0807/37/6/065103}.
\newblock \urlprefix\url{https://doi.org/10.1088/0143-0807/37/6/065103}

\bibitem{Nagle_1970}
Nagle, J.F., Bonner, J.C.: Numerical studies of the ising chain with long-range
  ferromagnetic interactions.
\newblock Journal of Physics C: Solid State Physics \textbf{3}(2), 352--366
  (1970).
\newblock \doi{10.1088/0022-3719/3/2/017}.
\newblock \urlprefix\url{https://doi.org/10.1088/0022-3719/3/2/017}

\bibitem{1990}
Wragg, M.J., Gehring, G.A.: The ising model with long-range ferromagnetic
  interactions \textbf{23}(11), 2157--2164 (1990).
\newblock \doi{10.1088/0305-4470/23/11/036}.
\newblock \urlprefix\url{https://doi.org/10.1088/0305-4470/23/11/036}

\bibitem{PhysRevB.52.3034}
Cannas, S.A.: One-dimensional ising model with long-range interactions: A
  renormalization-group treatment.
\newblock Phys. Rev. B \textbf{52}, 3034--3037 (1995).
\newblock \doi{10.1103/PhysRevB.52.3034}.
\newblock \urlprefix\url{https://link.aps.org/doi/10.1103/PhysRevB.52.3034}

\bibitem{Seth_2016}
Seth, S.: Combinatorial approach to exactly solve the 1d ising model.
\newblock European Journal of Physics \textbf{38}(1), 015104 (2016).
\newblock \doi{10.1088/1361-6404/38/1/015104}.
\newblock \urlprefix\url{https://doi.org/10.1088/1361-6404/38/1/015104}

\bibitem{Wang_2019}
Wang, W., D{\'{\i}}az-M{\'{e}}ndez, R., Capdevila, R.: Solving the
  one-dimensional ising chain via mathematical induction: an intuitive approach
  to the transfer matrix.
\newblock European Journal of Physics \textbf{40}(6), 065102 (2019).
\newblock \doi{10.1088/1361-6404/ab330c}.
\newblock \urlprefix\url{https://doi.org/10.1088/1361-6404/ab330c}

\bibitem{1978}
Baxter, R.J., Enting, I.G.: 399th solution of the ising model \textbf{11}(12),
  2463--2473 (1978).
\newblock \doi{10.1088/0305-4470/11/12/012}.
\newblock \urlprefix\url{https://doi.org/10.1088/0305-4470/11/12/012}

\bibitem{britanak2010discrete}
Britanak, V., Yip, P., Rao, K.: Discrete Cosine and Sine Transforms: General
  Properties, Fast Algorithms and Integer Approximations.
\newblock Elsevier Science (2010).
\newblock \urlprefix\url{https://books.google.co.in/books?id=iRlQHcK-r\_kC}

\end{thebibliography}

\end{document}